\documentclass[paper = a4,  
        pagesize = auto,    
        twoside=true,       
        titlepage = false,
        abstract = true,
]{scrartcl}

\usepackage[utf8]{inputenc}
\usepackage[T1]{fontenc}
\usepackage[english]{babel}
\usepackage{lmodern}
\usepackage[noblocks]{authblk}

\usepackage[style=authoryear,backend=biber,isbn=false,doi=true]{biblatex}
\usepackage{csquotes}
\usepackage{amsmath}
\allowdisplaybreaks[1]     
\usepackage{amssymb}
\usepackage{amsthm}
\usepackage{mathtools}

\usepackage{subfig}
\usepackage[english, ruled, vlined, linesnumbered]{algorithm2e}
\SetAlCapSty{textnormal}        
\usepackage{booktabs}
\usepackage{url}
\usepackage{siunitx}
\DeclareSIUnit[number-unit-product = {\,}]{\dt}{\Delta t}
\DeclareSIUnit[number-unit-product = {}]{\year}{yr}
\DeclareSIUnit[number-unit-product = {}]{\Phosphate}{P}
\DeclareSIUnit[number-unit-product = {}]{\Bracketo}{(}
\DeclareSIUnit[number-unit-product = {}]{\Bracketc}{)}
\usepackage{pdfpages}

\usepackage[
pdftitle={Unique steady annual cycle in marine ecosystem model simulations},%
pdfauthor={Markus Pfeil and Thomas Slawig}]{hyperref}

\usepackage{todonotes}

\title{Unique steady annual cycle in marine ecosystem model simulations}
\author[1]{Markus Pfeil}
\author[1]{Thomas Slawig}
\affil[1]{Kiel Marine Science (KMS) - Centre for Interdisciplinary Marine Science, Dep. of Computer Science, Kiel University, 24098 Kiel, Germany (\{mpf, ts\}@informatik.uni-kiel.de).}

\date{\vspace{-5ex}}

\begin{document}

  \maketitle

  \begin{abstract}
Marine ecosystem models are an important tool to assess the role of the ocean
biota in climate change and to identify relevant biogeochemical processes by
validating the model outputs against observational data. For the assessment of
the marine ecosystem models, the existence and uniqueness of an annual
periodic solution (i.e., a steady annual cycle) is desirable. To analyze the
uniqueness of a steady annual cycle, we performed a larger number of simulations
starting from different initial concentrations for a hierarchy of biogeochemical
models with an increasing complexity. The numerical results suggested that the
simulations finished always with the same steady annual cycle regardless of the
initial concentration. Due to numerical instabilities, some inadmissible
approximations of the steady annual cycle, however, occurred in some cases for
the three most complex biogeochemical models. Our numerical results indicate a
unique steady annual cycle for practical applications.
\end{abstract}

  \section{Introduction}
  \label{sec:Introduction}

    In the field of climate research, marine ecosystem models are an important tool
to assess the role of marine biogeochemical processes in climate change. These
models are generally an essential component for the analysis of biogeochemical
processes influencing the marine ecosystem. As part of the global carbon cycle,
the ocean, specifically, takes up $\textrm{CO}_{2}$ from the atmosphere and
stores it and, subsequently, the marine carbon affects the ocean biota. In
addition to the biogeochemical processes, a marine ecosystem model takes into
account the physical processes as well as the interplay between the physical and
biogeochemical processes wherefore a marine ecosystem model consists of a global
circulation model coupled to a biogeochemical model
\parencite[cf.][]{Fasham03, SarGru06, FenNeu04}. The physical processes
describing the ocean circulation are well known to include the equations and
variables. Conversely, there is generally no set of equations and variables
describing the biogeochemical processes. Therefore, many different
biogeochemical models with varying complexity exist which differ in the number
of state variables and parametrizations \parencite[e.g.,][]{KrKhOs10, KeOsEb12,
ISSMLN13, YoPoAn13, LBMAAB16}. For these different biogeochemical models,
validation and assessment is necessary to evaluate the simulated steady annual
cycle against observational data \parencite[cf.][]{FLSW01, SchOsc03}. This
requires a parameter optimization and a discussion of simulation results for
each biogeochemical model.

The biogeochemical model assessment is based on the existence and uniqueness of
periodic solutions. If different periodic solutions exist for one model
parameter vector, parameter optimizations or sensitivity studies are practically
impossible. Furthermore, the interpretation of numerically obtained steady
annual cycles is difficult if a non periodic solution of the continuous
equations exists. Therefore, theoretical results about the existence and
uniqueness of periodic solutions are helpful and desired to validate the
numerical results. However, the theoretical analysis is a challenging task so
far without a proof for a unique periodic annual cycle for complex
biogeochemical models \parencite{RosSla14, RosSla15}. In this paper, we analyzed
the uniqueness of a steady annual cycle by running a larger number of
simulations starting from different initial concentrations for a hierarchy of
biogeochemical models with an increasing complexity. In addition to the mostly
used constant initial concentration, we applied a wide variety of randomly
generated initial concentrations for the different biogeochemical models. A
unique steady annual cycle in these numerical simulations is, indeed, an
indication for practical applications but it does not in general allow any
conclusion about the existence and uniqueness of periodic solutions.

The paper is organized as follows: next, we describe marine ecosystem models
including the computation of steady annual cycles (Section \ref{sec:Model}). In
Section \ref{sec:InitialConcentrations}, we present a wide variety of different
initial concentrations. Numerical results of the steady annual cycle computation
starting from different initial concentrations are discussed in Section
\ref{sec:Results}. The paper closes with a summary and conclusions (Section
\ref{sec:Conclusions}).

  \section{Model description}
  \label{sec:Model}

    A marine ecosystem model describes the interplay of the ocean circulation and
the marine biogeochemical processes. The modeling of the marine ecosystem
comprises a given number of ecosystem species (or biogeochemical tracers), which
are substances in marine water and part of the biogeochemical cycle (i.e.,
subject to chemical and biochemical reactions). In a fully coupled marine
ecosystem model (also called \emph{online} model), the ocean circulation affects
the biogeochemical tracers and, vice versa, the biogeochemical tracers influence
the ocean circulation. Due to the full coupling, a simulation using such a model
is computationally expensive and, therefore, generally limited to single model
evaluations because a computation of a steady annual cycle requires a long-time
integration over several millennia \parencite{Osc06, BeDiWu08, Bryan84,
DaMcLa96}. In contrast to an online model, an \emph{offline} model neglects the
influence of the biogeochemical tracers on the ocean circulation. As a
result, this one-way coupling enables the application of a pre-computed ocean
circulation and reduces the computational effort. In this paper, we approximated
marine ecosystem models using an offline model with an increasing complexity of
the biogeochemical models introduced by \textcite{KrKhOs10, DuSoScSt05}.

\subsection{Model equations for marine ecosystems}
\label{sec:ModelEquation}

  A system of partial differential equations represents the marine ecosystem
  model. The number of modeled tracers determines the complexity of the marine
  ecosystem model and, thus, the size of the system of differential equations.
  In the rest of this paper, we consider marine ecosystem models using an
  offline model with $n_y \in \mathbb{N}$ tracers on a spatial domain
  $\Omega \subset \mathbb{R}^3$ (i.e., the ocean) and a time interval $[0,1]$
  (i.e., one model year). Function $y_i: \Omega \times [0,1] \rightarrow
  \mathbb{R}$, $i \in \left\{1, \ldots, n_y \right\}$, describes the tracer
  concentrations of tracer $y_i$ and $\mathbf{y} := \left( y_i
  \right)_{i=1}^{n_{y}}$ summarizes the tracer concentrations of all tracers.
  For $i = 1, \ldots, n_y$, the system of parabolic partial differential
  equations
  \begin{align}
  \label{eqn:Modelequation}
      \frac{\partial y_i}{\partial t} (x,t)
           + \left( D (x,t) + A(x,t) \right) y_i (x,t)
        &= q_i \left( x, t, \mathbf{y}, \mathbf{u} \right),
        & x \in \Omega, t &\in [0,1], \\
    \label{eqn:Boundarycondition}
      \frac{\partial y_i}{\partial n} (x,t) &= 0,
        & x \in \partial \Omega, t &\in [0,1],
  \end{align}
  describes the tracer transport of a marine ecosystem model. Here, the
  homogeneous Neumann boundary condition \eqref{eqn:Boundarycondition} includes
  the normal derivative and models no fluxes on the boundary.

  The ocean currents, modeled by spatially discretized advection and diffusion,
  transport the tracers in marine water. The linear operator $A: \Omega \times
  [0,1] \rightarrow \mathbb{R}$ describes the advection as
  \begin{align}
    \label{eqn:Advection}
    A(x,t) y_i (x,t) &:= \textrm{div} \left( v(x,t) y_i (x,t) \right),
    & x \in \Omega, t &\in [0,1],
  \end{align}
  $i \in \{1, \ldots, n_y\}$, using a given velocity field $v: \Omega \times
  [0,1] \rightarrow \mathbb{R}^3$. The diffusion operator $D: \Omega \times
  [0,1] \rightarrow \mathbb{R}$ models the turbulent effects of the ocean
  circulation but neglects the molecular diffusion of the tracers themselves
  because this is known to be much smaller than the diffusion induced by
  turbulence. Due to the quite different spatial scales in horizontal and
  vertical direction, the diffusion operator requires a splitting
  $D = D_h + D_v$ into a horizontal and a vertical part and an implicit
  treatment of the vertical part $D_v$ in the time-integration. Both directions
  are modeled in the second-order form as
  \begin{align}
    \label{eqn:Diffusion-horizontal}
    D_{h} (x,t) y_i (x,t) &:= -{\textrm{div}_h} \left( \kappa_h (x,t) \nabla_h
                                y_i (x,t) \right)
    & x \in \Omega, t &\in [0,1],\\
    \label{eqn:Diffusion-vertical}
    D_{v} (x,t) y_i (x,t) &:= -\frac{\partial}{\partial z}
            \left(\kappa_{v} (x,t)\frac{\partial y_i}{\partial z}(x,t)\right),
    & x \in \Omega, t &\in [0,1],
  \end{align}
  $i \in \{1, \ldots, n_y\}$, where $\textrm{div}_h$ and $\nabla_h$ denote the
  horizontal divergence and gradient, $\kappa_h, \kappa_v: \Omega \times [0,1]
  \rightarrow \mathbb{R}$ the diffusion coefficient fields and $z$ the vertical
  coordinate. The diffusion coefficient fields are identical for all tracers
  since the molecular diffusion is neglected.

  The biogeochemical model contains the biogeochemical processes modeled in the
  marine ecosystem. In addition to the biogeochemical model, the marine
  ecosystem model also takes the effects of the ocean dynamics into account and,
  therefore, contains the whole system \eqref{eqn:Modelequation} to
  \eqref{eqn:PeriodicCondition}. The nonlinear function $q_i: \Omega \times
  [0,1] \rightarrow~\mathbb{R}, \left( x, t \right) \mapsto q_i \left( x, t,
  \mathbf{y}, \mathbf{u} \right)$ represents the biogeochemical processes for
  tracer $y_i$, $i \in \{1, \ldots, n_y\}$. Indeed, these functions $q_i$
  depend, firstly, on space and time (for example, on the variability of the
  solar radiation), secondly, on the coupling to the other tracers and, thirdly,
  on $n_u \in \mathbb{R}$ model parameters $\mathbf{u} \in \mathbb{R}^{n_u}$
  (such as growth, loss and mortality rates or sinking speed). The
  biogeochemical model $\mathbf{q} = \left( q_i \right)_{i=1}^{n_{y}}$
  summarizes the biogeochemical processes of all tracers.

  An annually periodic solution of the marine ecosystem model (i.e., a steady
  annual cycle) fulfills in addition to \eqref{eqn:Modelequation} and
  \eqref{eqn:Boundarycondition}
  \begin{align}
   \label{eqn:PeriodicCondition}
    y_i (x, 0) &= y_i (x, 1), & x &\in \Omega,
  \end{align}
  for $i = 1, \ldots, n_y$. Therefore, we assume that the operators $A, D$ and
  the functions $q_i$ are also annually periodic in time.

\subsection{Biogeochemical models}
\label{sec:BiogeochemicalModels}

  The biogeochemical models differ in the given number of ecosystem species. In
  the present paper, we applied a hierarchy of five different biogeochemical
  models with an increasing complexity introduced by \textcite{KrKhOs10} as well
  as a biogeochemical model introduced by \textcite{DuSoScSt05}. In the
  following, we briefly introduce the biogeochemical models but we refer to
  \textcite{KrKhOs10, DuSoScSt05, PiwSla16} for a detailed description of the
  modeled processes and model equations. Table
  \ref{table:Parameter-Modelhierarchy} summarizes the model parameters of the
  biogeochemical models. Moreover, Table
  \ref{table:ParameterValues-Modelhierarchy} contains the assignment of the
  model parameters to the various biogeochemical models and the values used in
  this paper.

  \begin{table}[tb]
    \caption{Model parameters of the biogeochemical models.}
    \label{table:Parameter-Modelhierarchy}
    \centering
    \begin{tabular}{l l l}
      \hline
      Parameter & Description & Unit \\
      \hline
      $k_w$ & Attenuation coefficient of water & \si{\per \metre} \\
      $k_c$ & Attenuation coefficient of phytoplankton & \si{\Bracketo \milli \mol \Phosphate \per \cubic \metre \per \Bracketc \per \metre} \\
      $\mu_P$ & Maximum growth rate & \si{\per \day} \\
      $\mu_Z$ & Maximum grazing rate & \si{\per \day} \\
      $K_N$ & Half saturation constant for $\textrm{PO}_4$ uptake & \si{\milli \mol \Phosphate \per \cubic \metre} \\
      $K_P$ & Half saturation constant for grazing & \si{\milli \mol \Phosphate \per \cubic \metre} \\
      $K_I$ & Light intensity compensation & \si{\watt \per \square \metre} \\
      $\sigma_Z$ & Fraction of production remaining in $\textrm{Z}$ & \si{1} \\
      $\sigma_\text{DOP}$ & Fraction of losses assigned to $\textrm{DOP}$ & \si{1} \\
      $\lambda_P$ & Linear phytoplankton loss rate & \si{\per \day} \\
      $\kappa_P$ & Quadratic phytoplankton loss rate & \si{\Bracketo \milli \mol \Phosphate \per \cubic \metre \per \Bracketc \per \day} \\
      $\lambda_Z$ & Linear zooplankton loss rate & \si{\per \day} \\
      $\kappa_Z$ & Quadratic zooplankton loss rate & \si{\Bracketo \milli \mol \Phosphate \per \cubic \metre \per \Bracketc \per \day} \\
      $k_c$ & Attenuation coefficient of phytoplankton & \si{\Bracketo \milli \mol \Phosphate \per \cubic \metre \per \Bracketc \per \day} \\
      $\lambda'_P$ & Phytoplankton mortality rate & \si{\per \day} \\
      $\lambda'_Z$ & Zooplankton mortality rate & \si{\per \day} \\
      $\lambda'_D$ & Degradation rate & \si{\per \day} \\
      $\lambda'_\text{DOP}$ & Decay rate & \si{\per \year} \\
      $b$ & Implicit representation of sinking speed & \si{1}  \\
      $a_D$ & Depth-dependent increase of sinking speed & \si{\per \day} \\
      $b_D$ & Initial sinking speed & \si{\metre \per \day} \\
      \hline
    \end{tabular}
  \end{table}

  \begin{table}[tb]
    \caption{Assignment of the model parameters to the biogeochemical models and used parameter values. The model parameters of the MITgcm-PO4-DOP model correspond to those of the N-DOP model.}
    \label{table:ParameterValues-Modelhierarchy}
    \centering
    \begin{tabular}{l r r r r r}
      \hline
      Parameter & N & N-DOP & NP-DOP & NPZ-DOP & NPZD-DOP \\
      \hline
      $k_w$         & 0.02  & 0.02  & 0.02  & 0.02  & 0.02  \\
      $k_c$         &       &       & 0.48  & 0.48  & 0.48  \\
      $\mu_P$       & 2.0   & 2.0   & 2.0   & 2.0   & 2.0   \\
      $\mu_Z$       &       &       & 2.0   & 2.0   & 2.0   \\
      $K_N$         & 0.5   & 0.5   & 0.5   & 0.5   & 0.5   \\
      $K_P$         &       &       & 0.088 & 0.088 & 0.088 \\
      $K_I$         & 30.0  & 30.0  & 30.0  & 30.0  & 30.0  \\
      $\sigma_Z$    &       &       &       & 0.75  & 0.75  \\
      $\sigma_\text{DOP}$  & & 0.67 & 0.67  & 0.67  & 0.67  \\
      $\lambda_P$   &       &       & 0.04  & 0.04  & 0.04  \\
      $\kappa_P$    &       &       & 4.0   &       &       \\
      $\lambda_Z$   &       &       &       & 0.03  & 0.03  \\
      $\kappa_Z$    &       &       &       & 3.2   & 3.2   \\
      $\lambda'_P$  &       &       & 0.01  & 0.01  & 0.01  \\
      $\lambda'_Z$  &       &       &       & 0.01  & 0.01  \\
      $\lambda'_D$  &       &       &       &       & 0.05  \\
      $\lambda'_\text{DOP}$ & & 0.5 & 0.5   & 0.5   & 0.5   \\
      $b$           & 0.858 & 0.858 & 0.858 & 0.858 &       \\
      $a_D$         &       &       &       &       & 0.058 \\
      $b_D$         &       &       &       &       & 0.0   \\
      \hline
    \end{tabular}
  \end{table}

  Many biogeochemical processes depend on the amount of available light. Based
  on the astronomical formula of \textcite{PalPla76} and taking into account the
  ice cover, the exponential attenuation of water as well as phytoplankton (if
  included in the model), the light intensity is described by the light
  limitation function $I: \Omega \times [0, 1] \rightarrow \mathbb{R}_{\geq 0}$.
  Due to the decreasing light intensity, the ocean is divided into a euphotic
  (sun lit) zone of about \SI{100}{\metre} and an aphotic zone below. The
  biological production (for example, photosynthesis, grazing or mortality)
  takes place mainly in the euphotic zone, and particulate matter sinks to depth
  where it remineralizes according to the empirical law relationship
  \parencite{MKKB87}.
  
  The N model contains only one tracer modeling phosphate ($\textrm{PO}_4$) as
  inorganic nutrients (\textrm{N}) (i.e., $\mathbf{y} = \mathbf{y}_{\text{N}}$)
  and is the simplest biochemical model of the hierarchy
  \parencite[cf.][]{BacMai90, KrKhOs10}. Depending on available nutrients and
  light, the phytoplankton production (or biological uptake)
  \begin{align}
    \label{eqn:Phytoplankton}
    f_P: \Omega \times [0,1] \rightarrow \mathbb{R},
      f_P (x, t) &= \mu_P y_P^* \frac{I(x,t)}{K_I + I(x,t)}
                    \frac{\mathbf{y}_N (x,t)}{K_N + \mathbf{y}_N (x,t)}
  \end{align}
  is limited by a maximum production rate $\mu_P \in \mathbb{R}_{>0}$ and
  applies an implicitly prescribed concentration of phytoplankton $y_P^* =
  0.0028$~\si{\milli\mole\Phosphate\per\cubic\metre}. Table
  \ref{table:ParameterValues-Modelhierarchy} lists the $n_u = 5$ model
  parameters.

  The N-DOP model includes two tracers, nutrients (\textrm{N}) and dissolved
  organic phosphorus (\textrm{DOP}), i.e. $\mathbf{y} =
  (\mathbf{y}_{\textrm{N}}, \mathbf{y}_{\textrm{DOP}})$
  \parencite[cf.][]{BacMai91, PaFoBo05, KrKhOs10}. Using the same phytoplankton
  production \eqref{eqn:Phytoplankton} as the N model, the N-DOP contains
  $n_u = 7$ model parameters (see Table
  \ref{table:ParameterValues-Modelhierarchy}).

  The NP-DOP model comprises three tracers: nutrients (\textrm{N}),
  phytoplankton (\textrm{P}) and dissolved organic phosphorus (\textrm{DOP}),
  i.e., $\mathbf{y} = (\mathbf{y}_{\textrm{N}}, \mathbf{y}_{\textrm{P}},
  \mathbf{y}_{\textrm{DOP}})$ \parencite[cf.][]{KrKhOs10}. Instead of using an
  implicit treatment $y_P^*$ of phytoplankton, the NP-DOP model computes the
  phytoplankton production \eqref{eqn:Phytoplankton} using the explicit
  phytoplankton concentration $\mathbf{y}_{\text{P}}$. Using the implicitly
  prescribed zooplankton concentration $y_Z^* =
  0.01$~\si{\milli\mole\Phosphate\per\cubic\metre}, the zooplankton grazing
  \begin{align}
    \label{eqn:Zooplankton}
    f_Z: \Omega \times [0,1] \rightarrow \mathbb{R},
    f_Z (x,t) &= \mu_Z y_Z^* \frac{\mathbf{y}_P(x,t)^2}{K_P^2 + \mathbf{y}_P(x,t)^2}
  \end{align}
  models the loss of phytoplankton. Overall, this model includes the $n_u = 13$
  model parameters listed in Table \ref{table:ParameterValues-Modelhierarchy}.

  The NPZ-DOP model includes four tracers, nutrients (\textrm{N}), phytoplankton
  (\textrm{P}), zooplankton (\textrm{Z}) and dissolved organic phosphorus
  (\textrm{DOP}), i.e., $\mathbf{y} = (\mathbf{y}_{\textrm{N}},
  \mathbf{y}_{\textrm{P}}, \mathbf{y}_{\textrm{Z}}, \mathbf{y}_{\textrm{DOP}})$
  \parencite[cf.][]{KrKhOs10}. The phytoplankton production
  \eqref{eqn:Phytoplankton} and zooplankton grazing \eqref{eqn:Zooplankton} are
  the same as for the NP-DOP model but this model uses explicitly the
  zooplankton concentration $\mathbf{y}_{\text{Z}}$ instead of implicitly
  prescribed concentration $y_Z^*$. Table
  \ref{table:ParameterValues-Modelhierarchy} lists the $n_u = 16$ model
  parameters.
  
  The NPZD-DOP model comprises five tracers, nutrients (\textrm{N}),
  phytoplankton (\textrm{P}), zooplankton (\textrm{Z}), detritus (\textrm{D})
  and dissolved organic phosphorus (\textrm{DOP}), i.e., $\mathbf{y} =
  (\mathbf{y}_{\textrm{N}}, \mathbf{y}_{\textrm{P}},$ $\mathbf{y}_{\textrm{Z}},
  \mathbf{y}_{\textrm{D}}, \mathbf{y}_{\textrm{DOP}})$, and is the most complex
  biogeochemical model of the hierarchy \parencite[cf.][]{SOGES05, KrKhOs10}.
  Using the phytoplankton production \eqref{eqn:Phytoplankton} and zooplankton
  grazing \eqref{eqn:Zooplankton} as for the NPZ-DOP model, the NPZD-DOP
  contains $n_u = 18$ model parameters (see Table
  \ref{table:ParameterValues-Modelhierarchy}).

  The MITgcm-PO4-DOP model contains two tracers, phosphate ($\textrm{PO}_4$)
  and dissolved organic phosphorus (\textrm{DOP}), i.e. $\mathbf{y} = 
  (\mathbf{y}_{\textrm{N}}, \mathbf{y}_{\textrm{DOP}})$
  \parencite[cf.][]{DuSoScSt05}. This model resembles the N-DOP model and,
  therefore, we identified the $n_u = 7$ model parameters with those of the
  N-DOP model (see Table \ref{table:ParameterValues-Modelhierarchy}).

\subsection{Transport matrix method}
\label{sec:TransportMatrixMethod}

  The \emph{transport matrix method} (TMM) efficiently approximates the tracer
  transport of the ocean circulation by matrix-vector multiplications
  \parencite{KhViCa05, Kha07}. The discretized advection-diffusion equation can
  be written as a linear matrix equation because the application of the
  advection and diffusion operator, $A$ and $D$, on a spatially discretized
  tracer vector is linear. Therefore, the TMM replaces the direct
  implementation of a discretization scheme for the advection and diffusion by
  the application of matrices approximating the ocean circulation. In
  particular, the TMM approximates the ocean circulation only with these
  matrices including the influence of all parameterized processes represented
  in the underlying ocean circulation model on the transport
  \parencite{KhViCa05}.

  Using the TMM, each time step of the simulation for a marine ecosystem model
  consists only of two matrix-vector multiplications, modeling the ocean
  circulation, and an evaluation of the biogeochemical model. In order to
  discretize the advection-diffusion equation, we use a grid with
  $n_x \in \mathbb{N}$ grid points $\left( x_k \right)_{k=1}^{n_x}$ as spatial
  discretization of the domain $\Omega$ (i.e., the ocean) and the time steps
  $t_0, \ldots, t_{n_{t}} \in [0,1]$, $n_t \in \mathbb{N}$, specified by
  \begin{align*}
    t_j &:= j \Delta t, & j &= 0, \ldots, n_t, & \Delta t &:= \frac{1}{n_t},
  \end{align*}
  as an equidistant grid for the discretization of the time interval $[0,1]$
  (i.e., one model year). For time instant $t_j$, $j \in \{0, \ldots,
  n_t -1\}$, vector $\mathbf{y}_{ji} \approx \left( y_{i} \left( t_{j},
  x_{k} \right) \right)_{k=1}^{n_x} \in \mathbb{R}^{n_x}$, firstly, represents
  the numerical approximation of a spatially discrete tracer $y_i$, $i \in \{1,
  \ldots, n_x\}$, and $\mathbf{q}_{ji} \approx \left( q_i \left( x_k, t_j,
  \mathbf{y_j}, \mathbf{u} \right) \right)_{k=1}^{n_x} \in \mathbb{R}^{n_x}$,
  $\mathbf{u} \in \mathbb{R}^{n_u}$, the spatially discretized biogeochemical
  term $q_i$ for the tracer $y_i$. Using a reasonable concatenation,
  $\mathbf{y}_{j} := \left( \mathbf{y}_{ji} \right)_{i=1}^{n_y} \in
  \mathbb{R}^{n_y n_x}$ and $\mathbf{q}_j := \left( \mathbf{q}_{ji}
  \right)_{i=1}^{n_y} \in \mathbb{R}^{n_y n_x}$, secondly, combine the numerical
  approximations as well as the biogeochemical terms of all tracers at time
  instant $t_j$. With an explicit discretization of the advection and horizontal
  diffusion as well as an implicit discretization of the vertical diffusion, the
  application of a semi-discrete Euler scheme for \eqref{eqn:Modelequation}
  results in a time-stepping
  \begin{align*}
    \mathbf{y}_{j+1} &=  \left( \mathbf{I} + \Delta t \mathbf{A}_j
                            + \Delta t \mathbf{D}_j^h \right) \mathbf{y}_j
                     + \Delta t \mathbf{D}_j^v \mathbf{y}_{j+1}
                     + \Delta t \mathbf{q}_j \left( \mathbf{y}_j, \mathbf{u} \right),
        & j &= 0, \ldots, n_t -1,
  \end{align*}
  with the identity matrix $I \in \mathbb{R}^{n_x \times n_x}$ and the spatially
  discretized counterparts $\mathbf{A}_j, \mathbf{D}_j^h$ and $\mathbf{D}_j^v$
  of the operators $A, D_h$ and $D_v$ at time instant $t_j$, $j \in \{0, \ldots,
  n_t - 1\}$. Defining the explicit and implicit transport matrices
  \begin{align*}
    \mathbf{T}_{j}^{\text{exp}} &:= \mathbf{I} + \Delta t \mathbf{A}_j
                                    + \Delta t \mathbf{D}_j^h \in \mathbb{R}^{n_x \times n_x}, \\
    \mathbf{T}_{j}^{\text{imp}} &:= \left( \mathbf{I} - \Delta t \mathbf{D}_j^v
                                    \right)^{-1} \in \mathbb{R}^{n_x \times n_x}
  \end{align*}
  for each time instant $t_j$, $j \in \{0, \ldots, n_t - 1\}$, a time step of a
  marine ecosystem model simulation using the TMM is specified by
  \begin{align}
    \label{eqn:TMM}
    \mathbf{y}_{j+1} &= \mathbf{T}_{j}^{\text{imp}}
                     \left( \mathbf{T}_{j}^{\text{exp}} \mathbf{y}_j
                        + \Delta t \mathbf{q}_j \left( \mathbf{y}_j, \mathbf{u} \right)
                        \right)
                 =: \varphi_j \left( \mathbf{y}_j, \mathbf{u} \right),
        & j &= 0, \ldots, n_t - 1.
  \end{align}

  In practical computations, twelve sparse explicit and implicit transport
  matrices represent the monthly averaged tracer transport. These matrices are
  sparse because they are generated using a grid-point based ocean circulation
  model and the implicit ones (i.e., the inverse of the discretization matrices)
  include only the vertical part of the diffusion. For any time instant $t_j$,
  $j \in \{0, \ldots, n_t - 1\}$, the matrices are interpolated linearly. In
  the present paper, we used transport matrices computed with the MIT ocean
  model \parencite{MAHPH97} using a global configuration with a latitudinal and
  longitudinal resolution of \ang{2.8125} and \num{15} vertical layers
  \parencite[see][]{KhViCa05}.

\subsection{Computation of steady annual cycles}
\label{sec:ComputationSteadyAnnualCycles}

  For a marine ecosystem model, an annual periodic solution (i.e., a steady
  annual cycle) is in a fully discrete setting a fixed-point of the nonlinear
  mapping
  \begin{align*}
    \Phi &:= \varphi_{n_t -1} \circ \ldots \circ \varphi_{0}
  \end{align*}
  describing the time integration of \eqref{eqn:TMM} over one model year with
  $\varphi_j$, $j \in \{0, \ldots, n_t -1\}$, defined in \eqref{eqn:TMM}.
  Accordingly, an annual periodic solution fulfills 
  \begin{align*}
    \mathbf{y}_{n_t} &= \Phi \left( \mathbf{y}_0 \right) = \mathbf{y}_0
  \end{align*}
  applying the above iteration \eqref{eqn:TMM} over one model year. Starting
  from an arbitrary initial concentration vector $\mathbf{y}^{0} \in
  \mathbb{R}^{n_y n_x}$ and using the fixed model parameters $\mathbf{u} \in
  \mathbb{R}^{n_u}$, a classical fixed-point iteration takes the form
  \begin{align}
    \label{eqn:Spin-upIteration}
    \mathbf{y}^{\ell + 1} &= \Phi \left( \mathbf{y}^{\ell}, \mathbf{u} \right),
                          & \ell = 0, 1, \ldots.
  \end{align}
  If we interpret this fixed-point iteration \eqref{eqn:Spin-upIteration} as
  pseudo-time stepping or \emph{spin-up}, vector $\mathbf{y}^{\ell} \in
  \mathbb{R}^{n_y n_x}$ contains the tracer concentrations at the first time
  instant of model year $\ell \in \mathbb{N}$.

  The difference between two consecutive iterates determined by
  \begin{align}
    \label{eqn:StoppingCriterion}
    \varepsilon_{\ell} := \left\| \mathbf{y}^{\ell} - \mathbf{y}^{\ell - 1} \right\|_2
  \end{align}
  is a measure for the numerical convergence (i.e., the periodicity of the
  steady annual cycle) of the iteration \eqref{eqn:Spin-upIteration} for model
  year $\ell \in \mathbb{N}$.

  \section{Initial concentrations}
  \label{sec:InitialConcentrations}

    Each spin-up calculation starts with a given initial tracer concentration. The
default initial concentration is a constant global mean tracer concentration
where the nutrient tracer N is initialized with a global mean tracer
concentration of \SI{2.17}{\milli\mol\Phosphate\per\cubic\meter} and the other
tracers (i.e., the tracers DOP, P, Z and D) are initialized with a global mean
tracer concentration of \SI{0.0001}{\milli\mol\Phosphate\per\cubic\meter}.

We randomly generated initial tracer concentrations for each biogeochemical
model. Based on continuous uniform, normal and lognormal distribution
\parencite[e.g.,][]{Tho18}, we created \num{100} different initial
concentrations, respectively. For each tracer, the mean of the concentration
matched the value of the default initial concentration in each box of the
spatial discretization for every distribution, and all concentrations were
non-negative. Apart from that, the partitioning of the mass between the tracers
also corresponded to that of the default concentration, i.e., the major part of
the mass belonged to the tracer N while the mass contribution of the other
tracers in the overall mass was infinitesimal. The overall mass of each randomly
generated initial concentration, in particular, coincided with the overall mass
of the default concentration of the corresponding biogeochemical model. In
addition to the three various distributions, we randomly created \num{100}
initial concentrations for each biogeochemical model where the total
concentration of a tracer was present in a single box of the spatial
discretization, i.e., the concentration was zero in every box of the
discretization except one single box  containing the whole mass for this tracer.
The box containing the whole mass for a tracer was randomly determined so that
the boxes for the different tracers most likely differed. Analogous to the
distributions above, the mass partitioning between the tracers resembled that of
the default concentration.

We randomly varied the partitioning of the mass between the tracers for the
generation of initial concentrations. Using the three distributions (uniform,
normal and lognormal distribution) as well as using the concentration with the
whole concentration in a single box of the discretization, we created \num{100}
initial concentrations with respectively a random partitioning of the mass
between the tracers for each biogeochemical model except for the N model with
only one tracer. Moreover, we randomly generated \num{100} initial
concentrations similar to the default initial concentration, where the tracer
concentration of each tracer was globally constant, but the mass partitioning
between the tracers was randomly determined. Lastly, we used constant initial
concentrations with the same concentration for all tracers except for one
tracer. More specifically, we gradually increased the tracer concentration by
\SI{0.1}{\milli\mol\Phosphate\per\cubic\meter} starting from
\SI{0.0001}{\milli\mol\Phosphate\per\cubic\meter} (except for one tracer) and,
subsequently, reduced it for the remaining tracer (by number of tracers times
\SI{0.1}{\milli\mol\Phosphate\per\cubic\meter}) starting from
\SI{2.17}{\milli\mol\Phosphate\per\cubic\meter}.

In summary, we created \num{900} (and \num{400} for the N model) different
initial concentrations and up to \num{42} by the stepwise concentration
adjustment for each biogeochemical model.

  \section{Results}
  \label{sec:Results}

    The spin-up calculation resulted almost in the same approximation of the steady
annual cycle starting from different initial concentrations except for some
outliers. In this section, we present the numerical results obtained for a wide
variety of initial concentrations and assessed the accuracy of the calculated
approximations.

\subsection{Experimental setup}
\label{sec:ExperimentalSetup}

  For the calculation of a steady annual cycle, we used the marine ecosystem
  toolkit for optimization and simulation in 3D (Metos3D) \parencite{PiwSla16}
  and ran each spin-up over \num{10000} model years. We applied the parameter
  vectors listed in Table \ref{table:ParameterValues-Modelhierarchy} for the
  different biogeochemical models identifying the parameter vector of the N-DOP
  model with that of the MITgcm-PO4-DOP model.

  We assessed the approximations of the steady annual cycle using different
  initial concentrations based on the norm of difference
  \eqref{eqn:StoppingCriterion} and the accuracy of the approximation. For
  this purpose, we compared these approximations with a reference solution,
  denoted by $\mathbf{y}_{\text{default}}^{10000}$, namely the result obtained
  by a spin-up with Metos3D using the default initial concentration. We
  measured the accuracy of an approximation $\mathbf{x} \in
  \mathbb{R}^{n_y n_x}$ by the relative difference
  \begin{align}
    \label{eqn:relativeError}
    \frac{\left\| \mathbf{x} - \mathbf{y}_{\text{default}}^{10000} \right\|_2}
         {\left\| \mathbf{y}_{\text{default}}^{10000} \right\|_2}
  \end{align}
  and called this quantity \eqref{eqn:relativeError} the \emph{(relative)
  error} of the respective result $\mathbf{x}$.

\subsection{Numerical Results}
\label{sec:NumericalResults}

  \begin{figure}[!tb]
    \centering
    \subfloat[N model: Norm of difference \eqref{eqn:StoppingCriterion}.]{\includegraphics{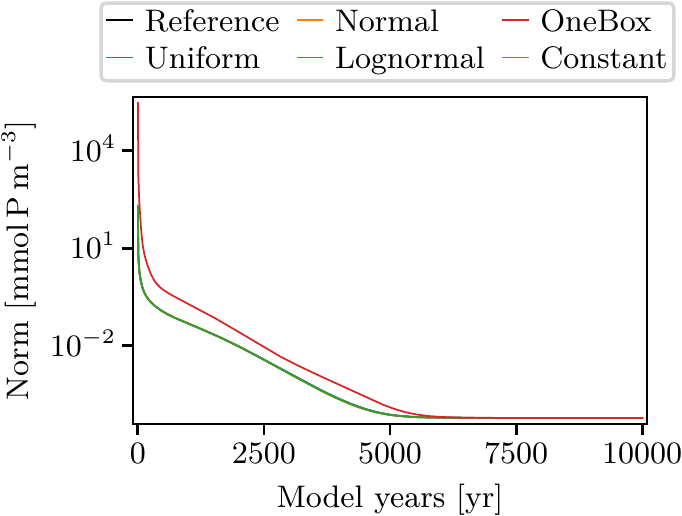}}
    \quad
    \subfloat[N model: Relative error \eqref{eqn:relativeError}.]{\includegraphics{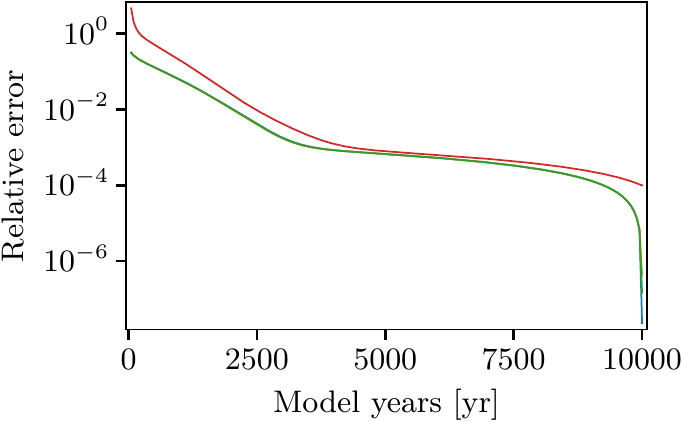}}
    \quad
    \subfloat[N-DOP model: Norm of difference \eqref{eqn:StoppingCriterion}.]{\includegraphics{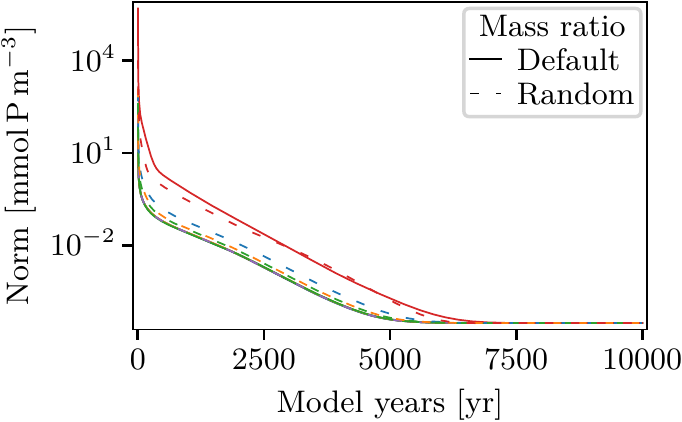}}
    \quad
    \subfloat[N-DOP model: Relative error \eqref{eqn:relativeError}.]{\includegraphics{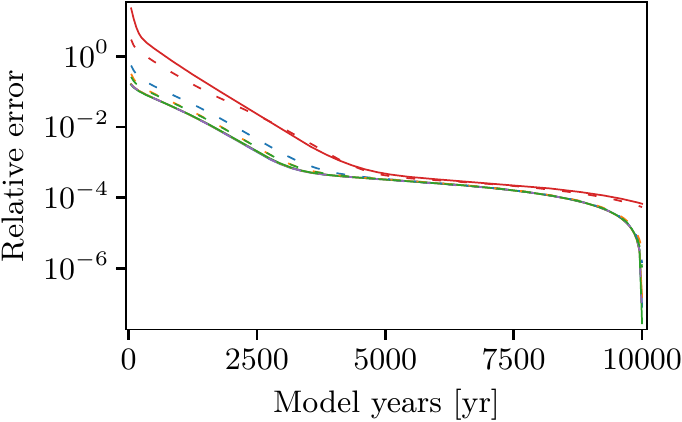}}
    \quad
    \subfloat[MITgcm-PO4-DOP model: Norm of difference \eqref{eqn:StoppingCriterion}.]{\includegraphics{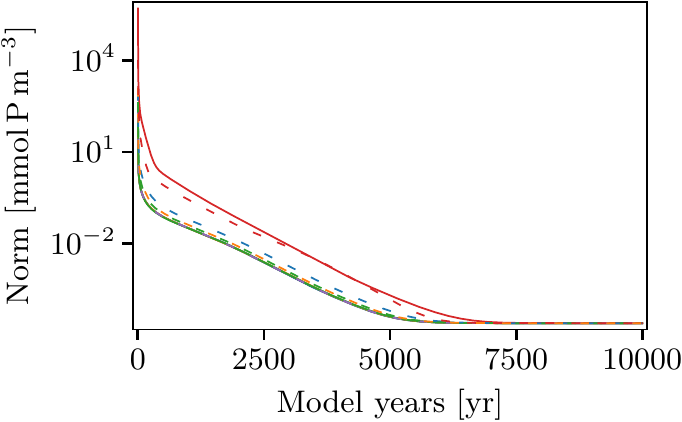}}
    \quad
    \subfloat[MITgcm-PO4-DOP model: Relative error \eqref{eqn:relativeError}.]{\includegraphics{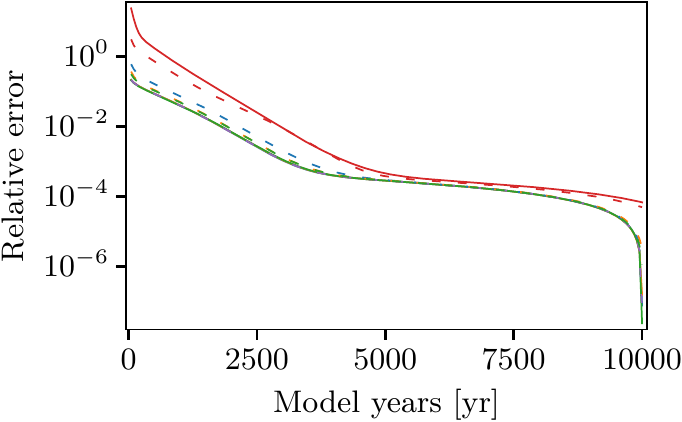}}
    \caption{Convergence of the spin-up using different initial concentrations
             for the N, N-DOP and MITgcm-PO4-DOP model. Shown are the norm of
             difference \eqref{eqn:StoppingCriterion} between consecutive
             iterations in the spin-up and the relative error
             \eqref{eqn:relativeError} for one exemplary parameter vector of
             each initial concentration type.}
    \label{fig:Convergence_1}
  \end{figure}

  \begin{figure}[!tb]
    \centering
    \subfloat[N model.]{\includegraphics{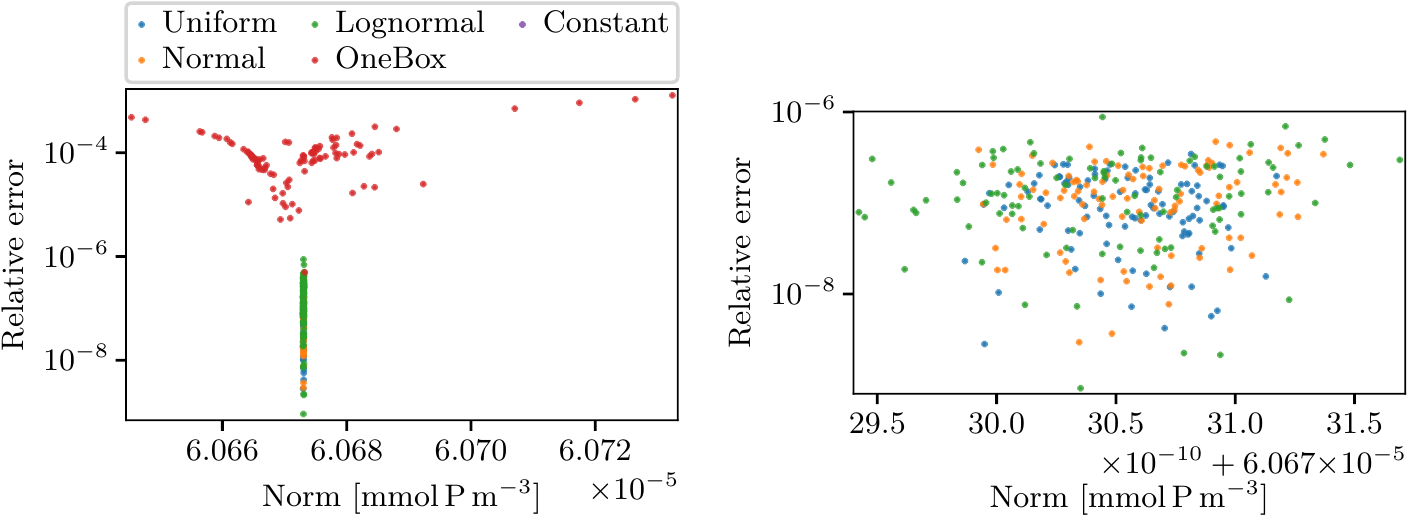}}
    \quad
    \subfloat[N-DOP model.]{\includegraphics{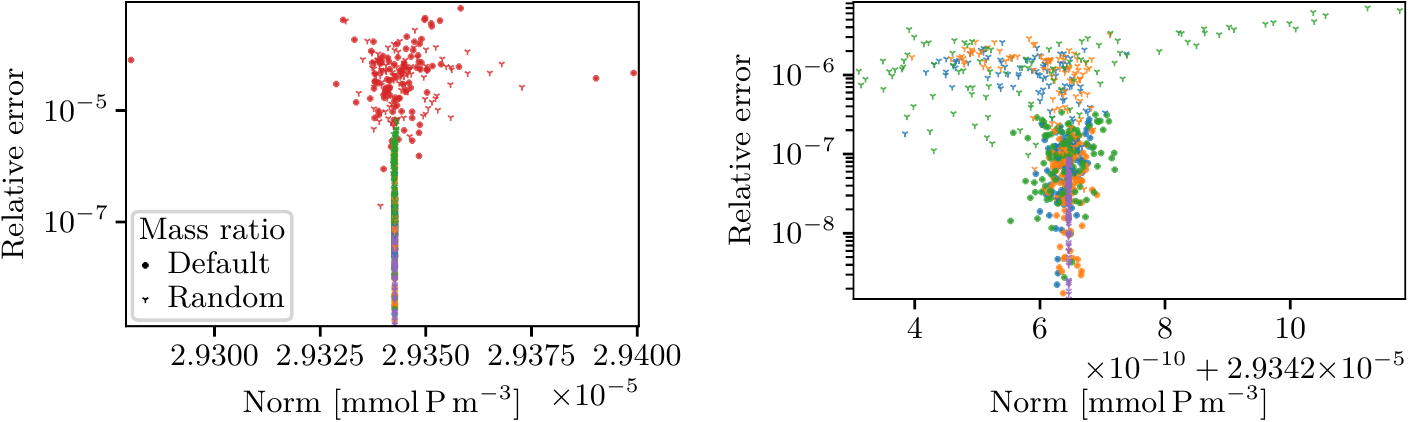}}
    \quad
    \subfloat[MITgcm-PO4-DOP model.]{\includegraphics{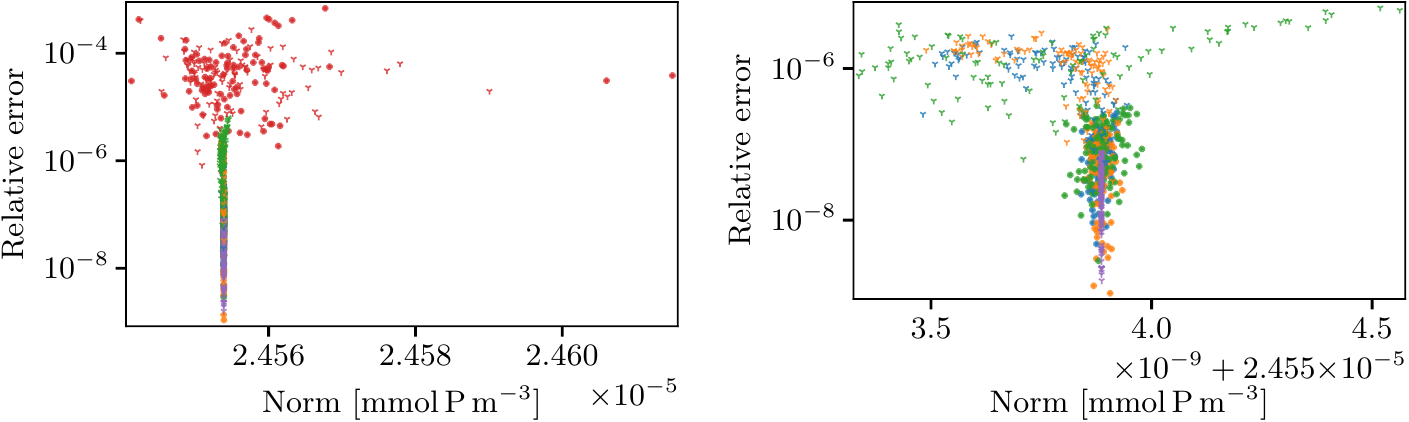}}
    \caption{Visualization of the norm of difference
             \eqref{eqn:StoppingCriterion} and the relative error
             \eqref{eqn:relativeError} for $\ell = 10000$ for the N, N-DOP and
             MITgcm-PO4-DOP model. Shown are the results for \num{100}
             different initial concentrations respectively of the various
             initial concentrations types. The figures in the right
             column contain a detail of the figure in the left column.}
    \label{fig:ScatterPlot_1}
  \end{figure}
  
  Regardless of the initial concentration, the spin-up calculation resulted in
  the same approximation of the steady annual cycle for the N, N-DOP and
  MITgcm-PO4-DOP model. Figure \ref{fig:Convergence_1} demonstrates a similar
  convergence behavior for all different initial concentrations indicating that
  the spin-up reached nearly the same accuracy of the norm of differences
  \eqref{eqn:StoppingCriterion}. Particularly, the spin-up ended with the same
  approximation of the steady annual cycle for the different initial
  concentrations (Figure \ref{fig:Convergence_1}, right column). Using an
  initial concentration with the whole concentration in only one box of the
  discretization for each tracer, the spin-up required several thousand model
  years to distribute the tracer concentration throughout the ocean.
  Consequently, the norm of differences is slightly larger and the accuracy
  after \num{10000} model years is slightly worse. To reach the accuracy of the
  other initial concentration types, further model years would be necessary.
  Furthermore, the marginal differences using a random partitioning of the mass
  (N-DOP and MITgcm-PO4-DOP model) resulted from the smaller concentration for
  the tracer N compared to the default initial concentration because most
  of the concentration was present as nutrients for the reference solution.
  Figure \ref{fig:ScatterPlot_1} shows the same results for all used initial
  concentrations. Except for the use of the initial concentration with the
  whole concentration in only one box of the spatial discretization for each
  tracer, the spin-ups reached almost the same norm of differences
  \eqref{eqn:StoppingCriterion}. More importantly, each of the spin-ups
  calculated an excellent approximation of the reference solution. Using the
  initial concentrations with the whole concentration in one single box for
  each tracer, the slightly larger relative error resulted from the required
  model years to distribute the tracer concentrations from the single boxes
  throughout the ocean. Nevertheless, these approximations adequately reflected
  the reference solution.

  \begin{figure}[!tb]
    \centering
    \subfloat[NP-DOP model: Norm of difference \eqref{eqn:StoppingCriterion}.]{\includegraphics{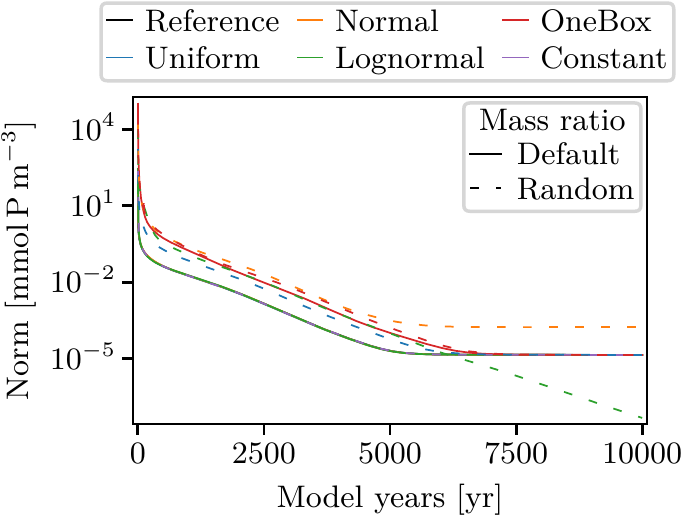}}
    \quad
    \subfloat[NP-DOP model: Relative error \eqref{eqn:relativeError}.]{\includegraphics{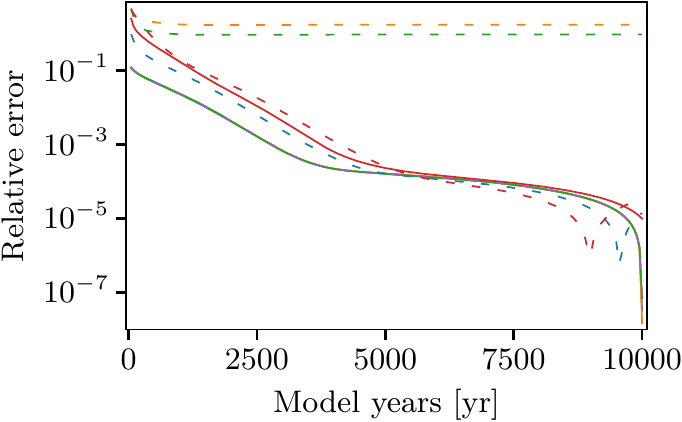}}
    \quad
    \subfloat[NPZ-DOP model: Norm of difference \eqref{eqn:StoppingCriterion}.]{\includegraphics{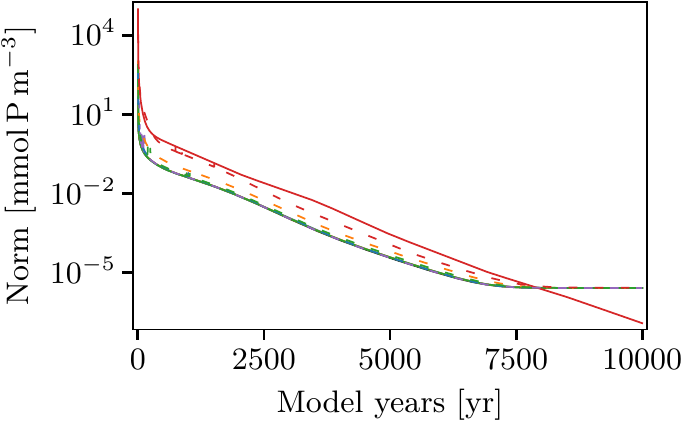}}
    \quad
    \subfloat[NPZ-DOP model: Relative error \eqref{eqn:relativeError}.]{\includegraphics{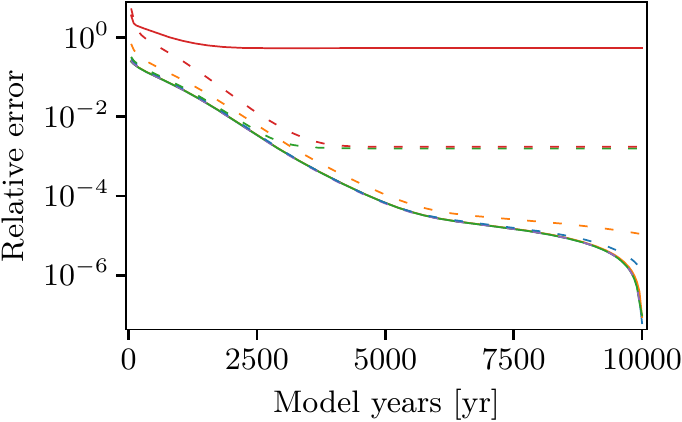}}
    \quad
    \subfloat[NPZD-DOP model: Norm of difference \eqref{eqn:StoppingCriterion}.]{\includegraphics{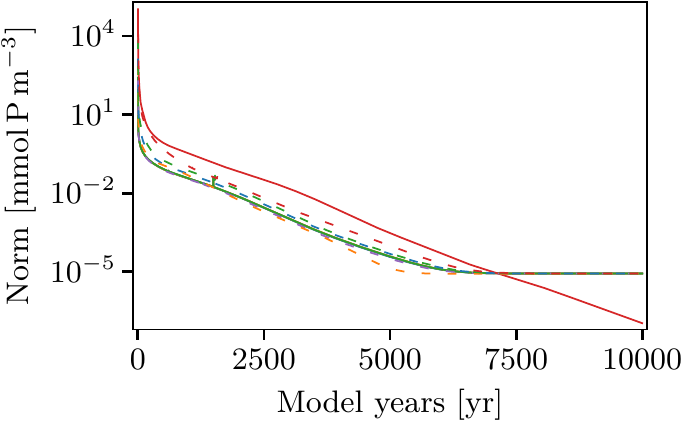}}
    \quad
    \subfloat[NPZD-DOP model: Relative error \eqref{eqn:relativeError}.]{\includegraphics{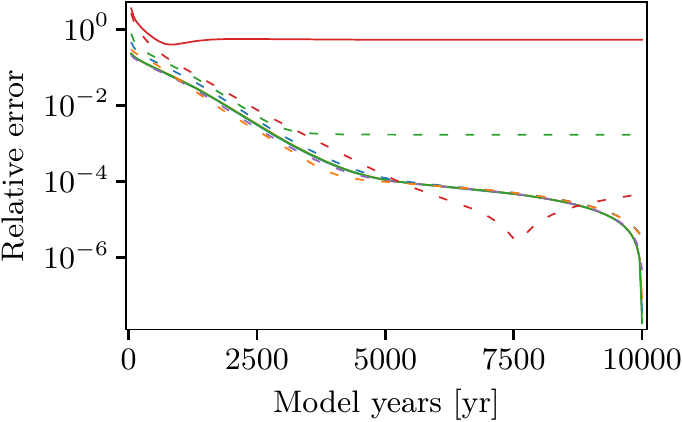}}
    \caption{Convergence of the spin-up using different initial concentrations
             for the NP-DOP, NPZ-DOP and NPZD-DOP model. Shown are the norm of
             difference \eqref{eqn:StoppingCriterion} between consecutive
             iterations in the spin-up and the relative error
             \eqref{eqn:relativeError} for one exemplary parameter vector of
             each initial concentration type.}
    \label{fig:Convergence_2}
  \end{figure}

  \begin{figure}[!tb]
    \centering
    \subfloat[NP-DOP model.]{\includegraphics{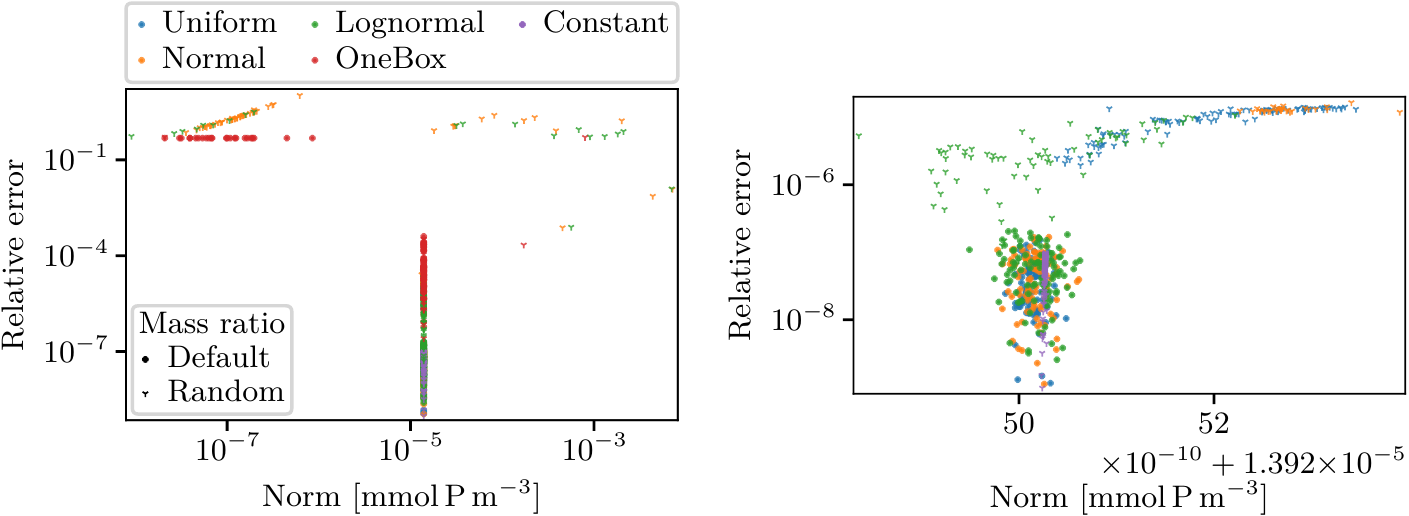}}
    \quad
    \subfloat[NPZ-DOP model.]{\includegraphics{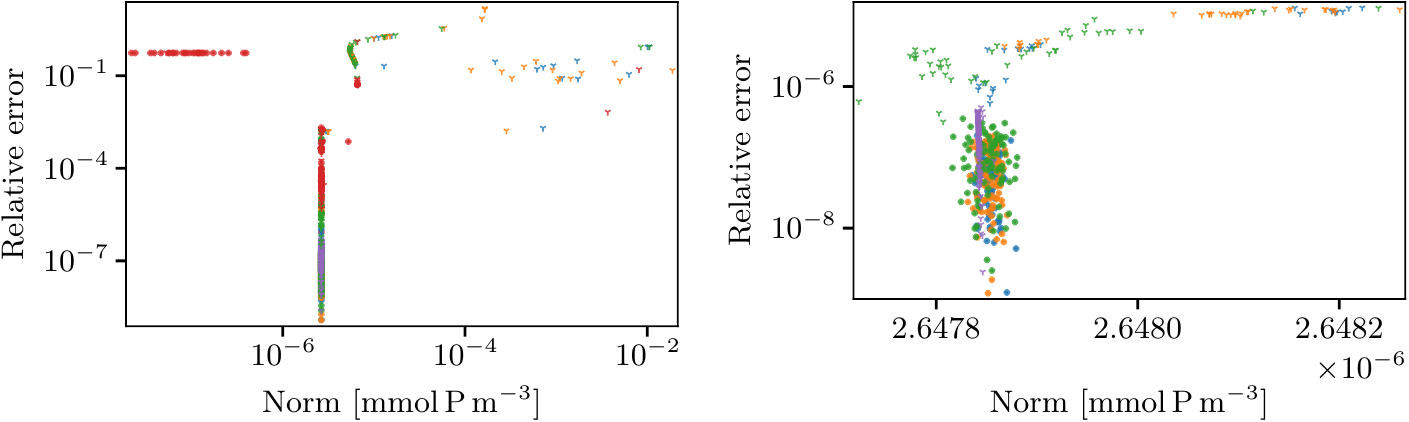}}
    \quad
    \subfloat[NPZD-DOP model.]{\includegraphics{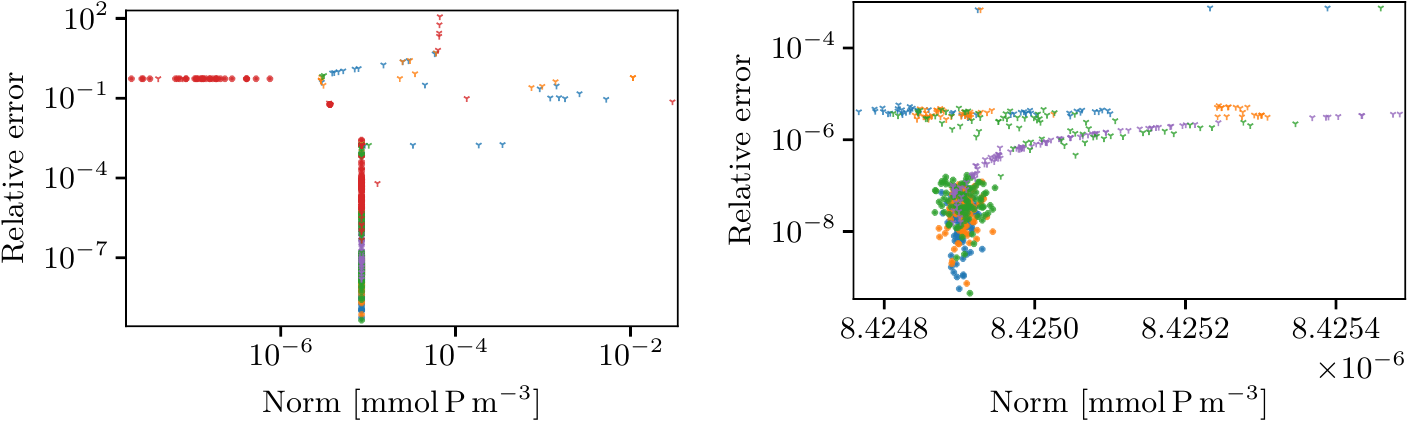}}
    \caption{Visualization of the norm of difference
             \eqref{eqn:StoppingCriterion} and the relative error
             \eqref{eqn:relativeError} for $\ell = 10000$ for the NP-DOP,
             NPZ-DOP and NPZD-DOP model. Shown are the results for \num{100}
             different initial concentrations respectively of the various
             initial concentration types. The figures in the right column
             contain a detail of the figure in the left column.}
    \label{fig:ScatterPlot_2}
  \end{figure}
  
  For the NP-DOP, NPZ-DOP and NPZD-DOP model, the initial concentration
  influenced the approximation of the steady annual cycle. The relative errors
  in Figure \ref{fig:Convergence_2} indicate the spin-up calculation of nearly
  the same approximation of the steady annual cycle using the different initial
  concentrations for each of the three biogeochemical models except for some
  outliers with a huge relative error. Clearly, the norm of differences
  \eqref{eqn:StoppingCriterion} was very small using the initial concentration
  that had been created with the lognormal distribution and random ratio of the
  tracer mass for the NP-DOP model as well as the initial concentration with the
  whole concentration in a single box for each tracer for the NPZ-DOP and
  NPZD-DOP model. Still, the spin-up ended with an invalid approximation of the
  steady annual cycle (Figure \ref{fig:Convergence_2}). In these approximations,
  first, all tracers were nearly constant, second, the tracer N contained more
  mass than was initially available and, third, the tracer concentrations of the
  other tracers (i.e., P, Z, D and DOP, if present) were exclusively negative.
  Likewise, the approximation computed with the initial concentration, which
  was generated with the normal distribution and random ratio of the tracer
  mass, was inadmissible because, as above, apart from the tracer N, all tracers
  were nearly constant and had exclusively negative concentrations (Figure
  \ref{fig:Convergence_2}). Conversely, the spin-up calculated a reasonable
  approximation of the reference solution for the NPZ-DOP and NPZD-DOP model
  starting from the initial concentration generated with the lognormal
  distribution and random ratio of the tracers (for the NPZ-DOP and NPZD-DOP
  model) or the initial concentration with the whole concentration in a single
  box and random ratio of the tracers (for the NPZ-DOP model), although the
  error (especially on the surface in the North Atlantic (Baffin Bay) or South
  Atlantic (southwest coast of Africa) for the tracers N and DOP) was slightly
  larger (Figure \ref{fig:Convergence_2}). Interestingly, the proportion of the
  tracer N compared to the other tracers was lowest of these three initial
  concentrations. Figure \ref{fig:ScatterPlot_2} summarizes the error of the
  approximations using the spin-up that was initialized with all different
  initial concentrations for the three different biogeochemical models NP-DOP,
  NPZ-DOP and NPZD-DOP. These approximations were mainly almost identical to the
  reference solution. As a result of the necessary distribution of the
  concentration throughout the ocean, the error was generally somewhat larger
  when an initial concentration with the whole concentration in single boxes
  was used. Especially in the case of using either a random ratio of the mass
  between the tracers or an initial concentration with the whole concentration
  in a single box for each tracer, there were, however, some initial
  concentrations for which the spin-up calculated inadmissible approximations
  containing negative tracer concentrations and, consequently, the error was
  large (Figure \ref{fig:ScatterPlot_2}).

  \section{Conclusions}
  \label{sec:Conclusions}

    The existence of a unique steady annual cycle is important for the validation
of marine ecosystem models. However, the theoretical analysis of the existence
and uniqueness of a periodic solution is a challenging task for the different
biogeochemical models \parencite[see e.g.,][]{RosSla14, RosSla15}. In this
study, we tested the uniqueness of a steady annual cycle by running a larger
number of simulations starting from different initial concentrations for
various biogeochemical models with increasing complexity. In addition to the
default constant initial concentration used for simulating marine ecosystem
models, we randomly generated a wide variety of different initial
concentrations.

Starting from different initial concentrations, the numerical simulations
finished with an almost identical approximation of a steady annual cycle,
except for some outliers for the most complex models. On the one hand, the
initial concentrations only marginally influenced the number of necessary model
years to obtain the steady annual cycle of the reference solution. For the
NP-DOP, NPZ-DOP and NPZD-DOP model, some initial concentrations, especially
those with random partitioning of the tracer mass, led, on the other hand, to
inadmissible steady annual cycles in which concentrations were negative.

The appearance of inadmissible steady annual cycles could be explained by a too
large step size of the Euler method. A too large step size could lead to a
draining of the biogeochemical model, which means the model could for instance
require more nutrients in a time step than were available in a box of the
discretization. Consequently, the concentration became negative. In fact, this
process depended on both the model parameters and the tracer concentrations
present in a box before the time step using the Euler method. Using an adequate
time step, the transport matrix method and the biogeochemical model did not
generate negative concentrations because the equations of the biogeochemical
model are quasi-positive \parencite[cf.][]{Pie10}. Hence, one cause of the
inadmissible steady annual cycles was a too large step size of the Euler method.
Interestingly, the inadmissible approximations occurred, firstly, only for the
three most complex biogeochemical models including quadratic terms (for the
grazing function and model equations) and, secondly, for initial concentrations
that had a random partitioning of tracer mass or the whole mass in a single box
for each tracer. In particular, the use of the initial concentration using the
whole mass in a box was unsuitable for the simulation in practice because
several millennia were initially needed for the distribution of the tracer
concentration throughout the ocean before the concentration approached the
steady annual cycle, which explains why a spin-up over a larger number of model
years was necessary \parencite[cf.][]{BeDiWu08, Bryan84, DaMcLa96}. In brief,
the inadmissible steady annual cycles resulted from numerical instabilities
during the simulation and did not indicate the existence of two different steady
annual cycles. Most notably, our numerical results suggested that the
simulations finished always with the same steady annual cycle regardless of the
initial concentration but this did not prove the existence and uniqueness of a
periodic solution. Future work should, therefore, include the theoretical
analysis of the existence and uniqueness of periodic solutions to marine
ecosystem models.

In summary, the main points of this paper are the following:
\begin{itemize}
  \item Regardless of the initial concentration, the spin-up calculation
        resulted in the same approximation for the N, N-DOP and MITgcm-PO4-DOP
        model.
  \item The initial concentration did not affect the approximation of the
        steady annual cycle for the NP-DOP, NPZ-DOP and NPZD-DOP model except
        for some outliers.
  \item For the NP-DOP, NPZ-DOP and NPZD-DOP model, the use of an initial
        concentration with a modified mass partitioning between the tracers
        could lead to an inadmissible approximation (occurrence of negative
        tracer concentrations).
\end{itemize}

  \section*{Code and data availability}
  
    The code used to generate the data in this publication  is available at
\url{https://github.com/slawig/bgc-ann} and \url{https://metos3d.github.io/}.
All used and generated data are available at
\url{https://doi.org/10.5281/zenodo.5644701} \parencite{PfeSla21aData}.

\printbibliography

\end{document}